\documentclass[preprint,aps]{revtex4}
\usepackage{graphicx}
\usepackage{float}
\usepackage{amsmath}    

\begin{document}
\title{Spiralicity and Motion on Cosmic Scales}
\author{E. Canessa\footnote{E-mail: canessae@ictp.it}}
\affiliation{{\it International Centre for Theoretical Physics, Trieste, Italy}}

\begin{abstract}
We hypothesize a simple scenario to associate spiral-shaped trajectories and asymptotic 
tangential velocities according to observations found on deep space scales.
As a difference with alternative Modified Newton Dynamics (MoND), in our scenario the 
dynamics at cosmic distances depends on the velocity of the rotating galaxy 
instead of its acceleration.  We assume dark matter to exert a quadratic fluid resistance 
force with an added non-linear drag force in the radial direction.
\vspace{1.0cm}


\end{abstract}
\maketitle

Spiral trajectories arise extensively in nature ranging from subatomic particles in 
a bubble chamber to huge galaxies in the universe.  Spiral galaxies are believed to 
form out of spinning gases and dark dust that made them to come together.  
The ultraviolet-light observations by the Hubble Space Telescope are unvealing a colourful
picture of the universe providing information about the birth and formation of distant galaxies.
Understanding the evolving universe at cosmic distance scales is a a multi-pronged pursuit.
It has profound impacts in physics areas ranging from astrophysics to cosmology.

In addition, it is known that the galaxy rotation curves approach an asymptotic velocity which 
has been studied in terms of modified gravity theories \cite{Bro06}, alternative Modified Newton
Dynamics (MoND) \cite{Mil83} and the presence of dark matter \cite{Van85}.  Thus the dynamics 
of flat spiral galaxies is still an open subject studied so far by modifying existing 
theories to fit observations.  In this work, we introduce some thoughts and speculations 
based on relativity theory to undertand the phenomenon of spiralicity at cosmic scales and 
to describe the circular motion and properties of visible, baryonic matter.

General relativity has recently been confirmed on the nearby universe \cite{Woj11}.
The predicted gravitational redshift of relativity has been found from the astronomical findings on
galactic clusters, where groups of thousands galaxies are held together by their own gravity field
to bend the light and to cause a change in the wavelength of the light. These observations may be
consistent with the existance of both dark matter and dark energy.  On the other hand, general relativity
still seems to hold \cite{Tys10} even if the so-called fifth force may enhance gravity at some
cosmic scales \cite{Jai13}.  Future cosmological experiments on the accelerating universe will require
data on millions of distant galaxies in order to reduce errors and verify predictions at deeper
space scales.  

The motivation for our side view is twofold.  First, our speculative work leads
to trace a physics scenario to associate asymptotic tangential speed curves for circular
motion with flat spiral-shaped trajectories.  Secondly, there is no need for
trial forms as for example the 1D $\mu (x)$ function used in MoND to explain the 
asympotic behaviour of velocity curves.  Our alternative formulation results in simple 
analytical dynamical functions in which the concept of dark matter is assumed 
to exert a quadratic fluid resistance force with an added non-linear drag force.  

The total energy of fast moving objects satisfies the key Einstein equation
$E =  m c^{2}$.  In terms of the momentun $p = mv = m_{o}v/\sqrt{1 - (v/c)^{2}}$, 
this relation can be written as 
\begin{equation}
E^{2} =  m_{o}^{2} c^{4} + c^{2} p^{2} \;\;\; , 
\end{equation}
with $m_{o}$ the object rest mass.  If $p$ is very large so that
$m_{o}^{2} c^{4} << c^{2} p^{2}$ (which means $v >> c/\sqrt{2}$), then the
derivative of the above energy-momentum formula can be approximated as 
\begin{equation}
dE = F dr \approx c\; dp = c \; d(mv)\;\;\; . 
\end{equation}
This amount of energy (work) may be seen as transferred by a force $F$ acting 
through a distance $r$ relating Newton's second law of motion $F = dp/dt = (dp/dr) v$,
with velocity $v = dr/dt$.  Thus, it follows a new type of Newton dynamics such that
\begin{equation}\label{eq:Ein} 
F \approx  (c/v) \; \frac{d}{dt}(mv) \;\;\; ,
\end{equation}
for velocities $\sim 0.7 c$.

MoND are theories that modifies Newton's second law 
of motion to explain unusual astronomical observations on spiral galaxies
\cite{Mil83}.  MoND assumes that the classical Newton's force law changes to the form
$F =\mu(a/a_{0})\; ma$, where $a_{0}$
is an acceleration constant and the function $\mu$ satisfies the
phenomenological conditions: $\mu \equiv 1$ for $a \gg a_{0}$, and
$\mu \equiv a/a_{0}$ for $a \ll a_{0}$.  Recent discussions
on the experimental possibility of a Newton's second law violation have concluded 
that this is feasible \cite{Ign07}.  However, as a difference with these phenomenological 
MoND theories that assume a given $\mu$ function to depend of the acceleration $dv/dt$, 
we can argue via eq.(\ref{eq:Ein}) that $\mu$ may rather depend on the 
(inverse of the) velocity $v$ of the moving object, {\it i.e.}, $F =\mu(c/v)\; ma$.

Let us introduce next our thoughts concerning spiralicity to help explaining the 
observed rotation curves of galaxies.  Based on eq.(\ref{eq:Ein}) we speculate the 
following new equations of motion in the radial direction 
\begin{equation}
F =  \mu (v_{_{\infty}}/v)\; \frac{d}{dt}(mv) \;\;\; , 
\end{equation}
This scenario for the fundamental Newton's force law on galactic scales
allows to embed dark matter imprints into MoND and to include effects of fluctuating 
mass in the universe.  One may consider either loss or gain of material in the process. 
For simplicity we shall analyse the motion of a constant mass traveling
through hypothetical fluid dark matter, such that
\begin{eqnarray}\label{eq:force}
\mu \equiv 1 \;\; \text{for small $v \ll v_{_{\infty}}$}  & \text{and} & F = m \left( \frac{dv}{dt} \right) \;\;\; , \nonumber \\
\mu \equiv  v_{_{\infty}}/v \;\; \text{for large $v \rightarrow v_{_{\infty}}$} & \text{and} & 
               F = mv\mu \left( \frac{dv}{dr} \right)  = mv_{_{\infty}} \left( \frac{dv}{dr} \right) \;\;\; . 
\end{eqnarray}
We shall show how this MoND-2 embraces the observed behaviour of flat rotation curves 
of turning galaxies \cite{Per96} through an hyphotetical non-linear drag force.
We adopt here the spiralicity concept first introduced by Hudge \cite{Hod92}.

Dark matter is believed to behave as a perfect fluid 
with neglected entropy changes and shear stress \cite{Arb06,Guz03,Peb99}.  
When dark matter is pictured as a fluid, then eq.(\ref{eq:force})
for motion in a gedanken "free fall" experiment satisfies 
\begin{equation}\label{eq:F}
F = m v_{_{\infty}} \left(\frac{dv}{dr}\right) = F_{g'} - \alpha v^{2} \;\;\; ,
\end{equation}
where $v \equiv v_{_{r}}(r)$ is the radial speed of the particle and
$F_{g'} = mg'$ is the gravitational force extended to cosmic scales; 
$g'$ corresponding to a centripetal acceleration which approaches $g$ for 
small distances \cite{San02}.  In MoND one relates the
gravitation acceleration $g$ to $\mu(g'/a_{0}){\bf g'}$ \cite{Lor09,San02}.
According to eq.(\ref{eq:F}) of MoND-2 the value of $g'$ is also constant.
Let us consider the ratio $g'/g = \lambda$, {\it i.e.}, to be proportional to the
acceleration of gravity such that $\lambda = 1$ (or $g' = g$ on Earth)
for small distances $r \ll r_{_{\infty}}$.  Using the $r_{_{\infty}}$ and $v_{_{\infty}}$
values given below to estimate the Hubble's law, we speculate MoND-2 to be valid for $g' \sim 10^{-10}g$.
In MoND the limit of low accelerations is of the order $10^{-10} \; ms^{-2}$ \cite{Lor09}.
The contribution of $F_{g'}$ on increasing $r$ to large scales becomes small because
$g' \ll g$.  From eq.(\ref{eq:vTrT}) we then have
\begin{equation}\label{eq:g}
g' = \left( \frac{ v_{_{\infty}}^{2} }{ r_{_{\infty}} } \right) = \lambda g \;\;\; . 
\end{equation}
It is interesting to note that
this relation and the Taylor series of eq.(\ref{eq:rapprox}) to second order in $\theta < 2\pi$
leads to the classic Newton dynamics for a free fall:
$r(t) = r_{_{\infty}} + v_{_{\infty}} t + (1/2) g t^{2}$
under the condition of small distances $v_{_{\infty}} t < r_{_{\infty}}$.
The latter giving a time constrain $t < 10^{20} / 10^{5} \; s \sim 10^{8} \; years$
(close to the age of the observable universe approximately equal to few billion years).
This time period is in correspondence with the maximum extension of
the spiral arm from its origin.

The last term on the right of eq.({\ref{eq:F}) is due to an opossite 
induced quadratic drag force $F_{d}$
as found, {\it e.g.}, from experiments of balls falling in water \cite{Mor00},
with $\alpha \rightarrow \frac{1}{2} C_{d}A\Omega$, and 
$\Omega$ being the fluid density and $C_{d}$ the drag coefficient of the resistance of
the object of area $A$ to the fluid environment.  The general solution of
eq.(\ref{eq:F}) for our idealized object falling with high velocity is found to be
\begin{equation}\label{eq:v}
v(r) = v_{_{\infty}} \tanh (r/r_{_{\infty}}) =  \sqrt{ \frac{mg'}{\alpha} } \tanh \left(\frac{\alpha}{m}\; r\right) \;\;\; ,
\end{equation}
with
\begin{equation}\label{eq:vTrT}
v_{_{\infty}} \equiv \sqrt{ mg'/\alpha }  \;\;\; \text{and} \;\;\; r_{_{\infty}}\equiv m/\alpha \;\;\; .
\end{equation}
where we identify $v_{_{\infty}}$ as an (asymptotic) terminal velocity.
This radial velocity for $r \ll r_{_{\infty}}$
rises linearly to the asymptotic value as a function of distance separation
\begin{equation}\label{eq:vlimit}
v(r) \sim \sqrt{ \frac{\alpha g'}{m} } = (v_{_{\infty}}/r_{_{\infty}}) \; r \;\;\; .
\end{equation}

For circular motion of an object around a center of rotation
the tangential velocity $v_{_{T}} \sim r$ is proportional to the distance 
from the center ({\it i.e.}, scale factor $r$) over the unit of time to cover 
an arc length with angle $\phi$.  This implies the centripetal acceleration term
$a_{_{centripital}} = v_{_{T}}^{2}/r$, with
\begin{equation}\label{eq:centr}
v_{_{T}}(r) = \omega \; r \;\;\; , 
\end{equation}
and $\omega = d\phi/dt$ the instantaneous angular velocity.
At this point let us then connect our free fall approach in the radial direction (driven 
by dark matter forces) to the traslational displacement of a turning, rigid object.
Let us extend eq.(\ref{eq:centr}), and using eq.(\ref{eq:vlimit}),
we approximate for the motion of any mass point on a rotating galaxy
\begin{equation}\label{eq:vT}
v_{_{T}}(r) = (r_{_{\infty}}/v_{_{\infty}})\; \omega \;v(r) \sim (\phi / 2\pi)\; v(r)  \;\;\; ,
\end{equation}
which tends to $r$ for small $r \ll r_{_{\infty}}$ and constant $\omega$.
We illustrated in Fig.1a the reduced function $(\phi / 2\pi)(v_{_{T}}(r)/v_{_{\infty}})$
versus the reduced positions $r/r_{_{\infty}}$.
It can be seen that in the velocity regime $v/v_{_{\infty}} \rightarrow 1$ (for cosmic distances),
our idealized traslational velocity curves in eq.(\ref{eq:vT}) approach an asymptotic velocity limit
because the hyperbolic function in $v$ tends to unity.  This behaviour is in reasonable 
agreement with observations \cite{Per96,Rub80}.

A crude estimate of the $\phi$-angle may follow from Hubble's law
$V = H_{0}D$, where $H_{0} \sim 2.5 \times 10^{-18} \; s^{-1}$ is the 
Hubble parameter relating (the redshift $z$ and) the distance measurement
$D$ to a given galaxy and its recession velocity $V$.  For a 
galaxy of effective radius of rotation $r_{_{\infty}} \sim 10^{20} \; m$ and 
rotation velocity $v_{_{\infty}} \sim 100 \; km/s$, we obtain $\phi \sim 1'$.
Embedded in this relation is the scale factor of the universe, suggesting some
kind of correlation between galactic dynamics with the expansion of the universe.

By eqs.(\ref{eq:F}), (\ref{eq:v}) and (\ref{eq:g}), it
is strightforward to show that the acceleration in MoND-2 satisfies
$a^{'} = \mu (v/v_{_{\infty}}) a = (v_{_{\infty}}/v) a$, where $a=dv/dt$
and the net external force becomes
\begin{equation}\label{eq:F1}
F = \frac{m}{r_{_{\infty}}} [v_{_{\infty}}^{2} - v^{2}] =
     \frac{mv_{_{\infty}}^{2}}{r_{_{\infty}}} [1 -  \tanh^{2}(r/r_{_{\infty}})] =
    F_{g'}\;  \text{sech}^{2}\left(\frac{\alpha}{m}\; r\right) \;\;\; ,
\end{equation}
which for small $r$, and in the absence of no other drag force, it approaches the 
classical force of gravity $F_{g} = mg$ (in the radial direction) as if objects were 
in free fall without fluid dark matter influences.  Crearly the above relation is not 
valid in the terminal upper limit $v = v_{_{\infty}} = c$.

There exists a force of gravitational attraction between two objects that reaches 
out over huge distances.  Then the net force $F$ matches Newton's gravitational force 
law between the falling object of mass $m = m_{N}$ and the rest of 
the galaxy mass $M = \sum_{i=1}^{N-1} m_{i}$.
Using eqs.(\ref{eq:F1}) and (\ref{eq:g}), in conjuction with the limiting behviour 
of $v(r\ll r_{_{\infty}})$ in eq.(\ref{eq:vlimit}), it follows that
\begin{equation}
\frac{mMG}{r^{2}} = \frac{mv^{2}}{r_{_{\infty}}} \left[ \left( \frac{v_{_{\infty}}}{v} \right)^{2} - 1 \right]  \;\;\;  ,
\end{equation}
or
\begin{equation}\label{eq:Mv4}
MGg' = v^{4} \left[ \left( \frac{v_{_{\infty}}}{v} \right)^{2} - 1 \right]  \;\;\; .
\end{equation}
Hence for $v \ll v_{_{\infty}}/\sqrt{2} = mg'/2\alpha$ and eq.(\ref{eq:vT}), 
a Tully-Fisher-like (TF) relation
$M \sim v_{_{T}}^{4}$ is satisfied consistenly with MoND \cite{Mil83,San02}.
Notwithstanding there are doubts about the necessity of halo dark matter around galaxies
to explain this TF relation based on the property of spacetime \cite{Car00}.
There is yet another possibility to explain the hidden matter in terms of a purely 
dynamical effect due to the curvature of spacetime \cite{Riz10}.

Let us account next for the trajectory curves $r(t) = (2\pi /\phi) r_{_{T}}(t)$ 
obtained within MoND-2.  So far 
we have shown that the asymptotic behaviour of our object's velocity $v = dr/dt$ is 
driven by the hyperbolic tangent function in eq.(\ref{eq:v}).  Integrating this equation 
we found that $r$ satisfies
\begin{equation}\label{eq:sinhr}
\sinh (r/r_{_{\infty}}) = \exp (v_{_{\infty}}t/r_{_{\infty}})  \;\;\; .
\end{equation}
For increasing $r \rightarrow r_{_{\infty}}$, this relation becomes
\begin{equation}\label{eq:rapprox}
r(t) \approx r_{_{\infty}} \exp (v_{_{\infty}}t/r_{_{\infty}})  \;\;\; .
\end{equation}
Setting $\theta / 2\pi = v_{_{\infty}}t/r_{_{\infty}}$, 
it can be rewritten in parametric form as
\begin{eqnarray}\label{eq:parametric}
x = r \cos \theta & = & r_{_{\infty}} \; e^{\theta / 2\pi} \cos \theta   \;\;\; ,  \nonumber \\ 
y = r \sin \theta & = & r_{_{\infty}} \; e^{\theta / 2\pi} \sin \theta   \;\;\; .
\end{eqnarray}
This corresponds to a flat logarithmic spiral as illustrated in Fig.1(b) for a complete 
revolution ($0 <\theta < 2\pi$).  In this regard, the spiral structure of the Via Lactea is 
found to display for example logarithmic arms of the form in eq.(\ref{eq:parametric}) \cite{Val02}.

We have seen that the alternative MoND-2 leads to associate the asymptotic $v(r)$ curves of 
eq.(\ref{eq:v}) or eq.(\ref{eq:vT}) with the flat spiral-shaped trajectories
$r(t)$ of eq.(\ref{eq:parametric}).  
Astronomical data show that only a small percentage of the universe is made
of luminous matter described by the Standard Model of fundamental particle physics.
The rest of the universe may consist of dark matter and dark energy \cite{Lor09}.  
In our MoND-2 the phenomenological MoND function $\mu$ of Milgrom \cite{Mil83} 
depends on the velocity of the object instead of its acceleration.
Furthermore, the asympotic behaviour of velocity curves within 
MoND-2 does not depends on arbitrary forms for a $\mu (x)$. 
More broadly, we think that the alternative force law in eq.(\ref{eq:force}) could be 
useful to understand how huge (Bullet) clusters of galaxies interact and influence each other.

Dark matter fluid seems to behave sufficiently classically on galaxy scales \cite{Bal04}.
It is usually considered as a system of collisionless particles and, consequently, 
it leads to an attractive gravitational effect like the mass of stars, planets, etc.  
If fluid dark matter permeates all the universe, flowing around and between galaxies,
then dark matter may also pose fluid resistance on the motion of visible baryonic  
matter on cosmic scales.  It may exert its influence on normal matter through 
an induced drag force $F_{d} \propto v^{2}$ acting on the moving objects
(through the curved space on galactic scales \cite{Can07}).  
The linear drag force proportional to the velocity $-\beta v$ has been previoulsy 
considered in \cite{Ume94} together with a dark matter potential to study 
the formation of condensed objects in the early universe.

A mass-energy relation follows from $dE = F dr$ and the force law eq.(\ref{eq:F1}).  
For a moving object through the fluid dark matter we find
\begin{equation}\label{eq:E}
E = F_{g'} \int  \text{sech}^{2} (r/r_{_{\infty}}) dr =
  m v_{_{\infty}} v = \sqrt{ \frac{mg'}{\alpha} } \; mv  \;\;\; ,
\end{equation}
which is potential energy $E = mg' r = F_{g'}r$ in the limit $r\ll r_{_{\infty}}$
({\it c.f.}, eq.(\ref{eq:vlimit})).  On the other limit, $v \rightarrow v_{_{\infty}}$, it 
approaches the value 
\begin{equation}\label{eq:Elimit}
E = m v_{_{\infty}}^{2}   \;\;\; .
\end{equation}
A MoND-2 energy density $\rho = E/Ar$ may relate dark energy and the accelerating 
expansion of space via the equation of state 
$w(z) = P/\rho$, with the dark energy pressure $P$ relating
our MoND-2 force expression in eq.(\ref{eq:F1}).
However, the idea that dark matter particles form a fluid ({\it e.g.}, a quantum 
fluid \cite{Bal04,Mur10}) still requires to figure out how this fluid was created 
\cite{Ume94} and how much energy is tied up.
In our scenario of spiralicity, the acceleration $g$ has been modified by a constant
to large distances and a new drag term has also been introduced and attributed to 
dark matter, contrary to the 
whole purpose of previous phenomenological modified gravity models
adopted to eliminate the need for dark matter \cite{Mil83}. 
While not wishing to distract those engrossed in these more rigourous formalisms, our hope here has been
to open another front of research for the dynamics of beautiful spiral patterns encoded at cosmic scales.

\newpage

\begin{figure}[ht!]
  \begin{center}
      \includegraphics[width=16.0cm]{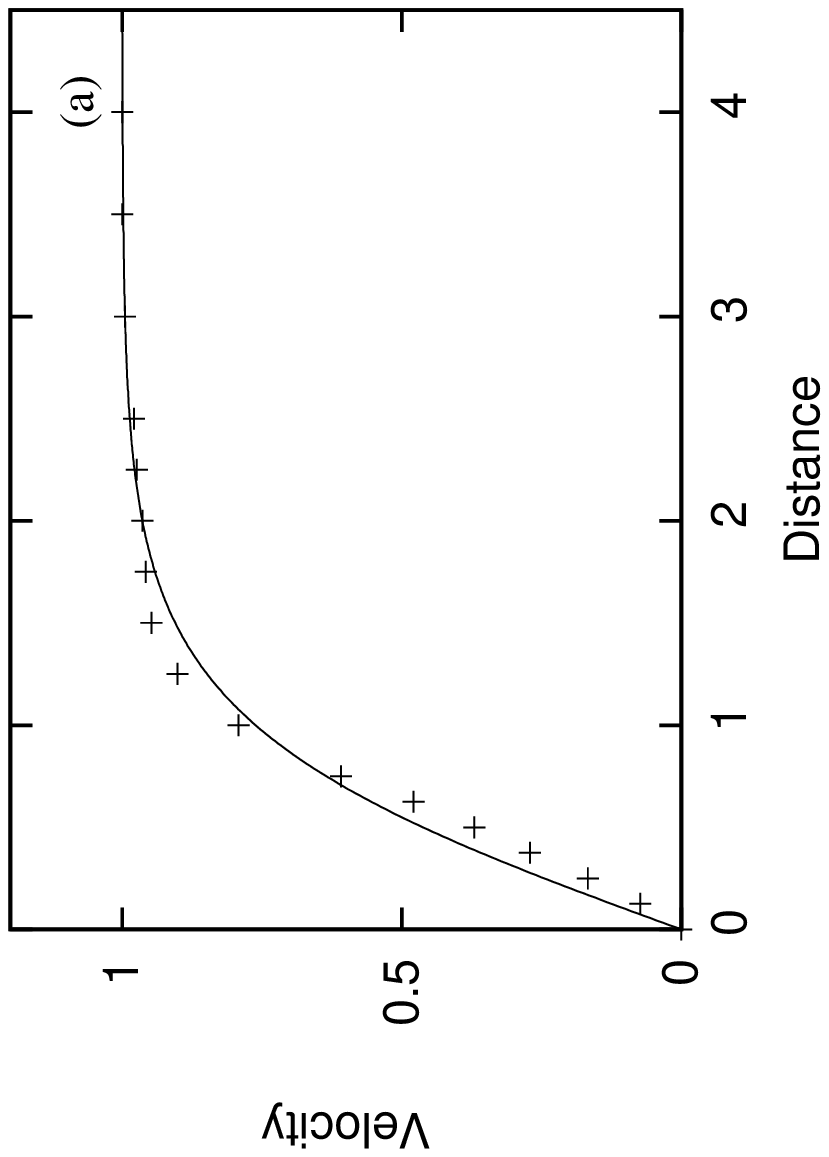}
  \end{center}
\end{figure}

\newpage

\begin{figure}[ht!]
  \begin{center}
      \includegraphics[width=16.0cm]{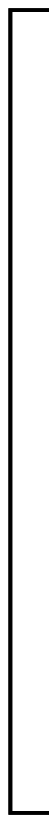}
  \end{center}
\end{figure}

\begin{figure}[ht!]
  \caption{Onset of spiralicity within MoND-2: (a) asymptotic, reduced tangential velocity curves for 
the circular motion of eq.(\ref{eq:parametric}) and (b) associated flat spiral-shaped trajectories 
from eq.(\ref{eq:vlimit}) for $0 <\theta /2\pi = (v_{_{\infty}}/r_{_{\infty}})t < 1$, and
times $t < \sim 10^{8} \; years$ smaller than the age of the universe. Experimental 
velocity data from \cite{Rub80} (crosess) reduced by a factor of $4 \; km/s$ of plane galaxy NGC3672 
versus distance from the nucleus reduced by $192 \; kpc$.}
\end{figure}


\begin{thebibliography}{99}

\bibitem{Bro06} J.R. Brownstein and J.W. Moffat, Astrophys. J. {\bf 636} (2006) 721; preprint ArXiv: 0506370 [astro-ph].
\bibitem{Mil83} M. Milgrom, Astrophys. J. {\bf 270} (1983) 365; preprint ArXiv: 0801.3133v2 [astro-ph].
M. Milgrom, Astrophys. J. {\bf 270} (1983) 365; preprint ArXiv: 0801.3133v2 [astro-ph].
\bibitem{Van85} T.S. Albada, J.N. Bahcall, K. Begeman and R. Sancisi, Astrophys. J. {\bf 295} (1985) 305.
\bibitem{Woj11} R. Wojtak, S.H. Hansen and J. Hjorth, Nature {\bf 477} (2011) 567.
\bibitem{Tys10} J.A. Tyson, Nature {\bf 464} (2010) 172.
\bibitem{Jai13} B. Jain, V. Vikram and J. Sakstein, Astrophysical J. {\bf 779} (2013) 39.
\bibitem{Ign07} A. Yu. Ignatiev, Phys. Rev. Lett. {\bf 98} (2007) 101101; Phys. Rev. D 
{\bf 77} (2008) 102001.
\bibitem{Per96} M. Persic, P. Salucci and F. Stel, Mon. Not. R. astr. Soc., {\bf 281} (1996) 27.
\bibitem{Hod92} P.W. Hodge in {\it "The Andromeda Galaxy"}, (Kluwer Academic Publisher, The Netherlands, 1992).
\bibitem{Arb06} A. Arbey, Phys. Rev. D {\bf 74} (2006) 043516.
\bibitem{Guz03} F.S. Guzman {\it et al.}, Rev. Mex. Fis. {\bf 49} (2003) 203.
\bibitem{Peb99} P.J.E. Peebles and A. Vilenkin, Phys. Rev. D {\bf 60} (1999) 103506.
\bibitem{San02} R.H. Sanders and S.S. McGaugh, ARA\&A {\bf 40} (2002) 263.
\bibitem{Lor09} V.A. De Lorenci, M. Fa\'undez-Abans and J.P. Pereira, A\&A {\bf 503} (2009) L1.
\bibitem{Mor00} M. Mordant and J.-F. Pinton, Eur. Phys. J {\bf B18} (2000) 343.
\bibitem{Rub80} V.C. Rubin  {\it et al.}, AJ {\bf 238} (1980) 471.
\bibitem{Car00} M. Carmeli, Int. J. Theor. Phys. {\bf 39} (2000) 1397.
\bibitem{Riz10} L. Rizzi {\it et al.}, Phys. Rev. D {\bf 82} (1010) 027301.
\bibitem{Val02} J.P. Vall\'ee, Astrophys. J. {\bf 566} (2002) 261.
\bibitem{Bal04} E.A. Baltz, SLAC Summer Institute on Particle Physics (SSI04), Aug. 2004
\bibitem{Can07} E. Canessa, Physica A {\bf 375} (2007) 123; {\it ibid} {\bf 385} (2007) 185.
\bibitem{Ume94} M. Umemura and J. Fukue, PASJ {\bf 46} (1994) 567.
\bibitem{Mur10} C. Lee, review on H. Murayama idea for a quantum fluid universe; \newline 
     http://arstechnica.com/science/news/2010/01/quantum-universe.ars

\end{thebibliography}
\end{document}